\documentclass[aps,prl,twocolumn,superscriptaddress]{revtex4-2}

\usepackage[utf8]{inputenc}
\usepackage[T1]{fontenc}
\usepackage{amsmath}
\usepackage{amssymb}
\usepackage[margin=2cm]{geometry}
\usepackage[dvipsnames]{xcolor}
\usepackage{graphicx}
\usepackage[sort&compress]{natbib}
\usepackage{changes}
\usepackage{tikz}
\usetikzlibrary{arrows,shapes}
\usetikzlibrary{shapes.multipart}
\usepackage[compat=1.1.0]{tikz-feynman}
\usepackage{contour}
\usepackage{feynmp}
\usepackage{tikzpagenodes}

\newcounter{comment}

\newcommand{\HMcolor}{red}

\newcommand{\MDcolor}{BrickRed}

\newcommand{\CMcolor}{orange}

\definecolor{DarkGreen}{RGB}{0,105,62}
\newcommand{\JMMCcolor}{DarkGreen}

\newcommand{\PPcolor}{blue}

\newcommand{\VBcolor}{brown}

\newcommand{\FSBcolor}{violet}

\newcommand{\JScolor}{purple}

{\refstepcounter{comment}%
\begin{quote}
\color{\JMMCcolor}\small$\blacksquare$ \textbf{\underline{Comment} $\sharp$\thecomment.~JMMC:}}%
{\end{quote}}

{\refstepcounter{comment}%
\begin{quote}
\color{\HMcolor}\small$\blacksquare$ \textbf{\underline{Comment} $\sharp$\thecomment.~HM:}}%
{\end{quote}}

{\refstepcounter{comment}%
\begin{quote}
\color{\MDcolor}\small$\blacksquare$ \textbf{\underline{Comment} $\sharp$\thecomment.~MD:}}%
{\end{quote}}

{\refstepcounter{comment}%
\begin{quote}
\color{\CMcolor}\small$\blacksquare$ \textbf{\underline{Comment} $\sharp$\thecomment.~CM:}}%
{\end{quote}}

{\refstepcounter{comment}%
\begin{quote}
\color{\PPcolor}\small$\blacksquare$ \textbf{\underline{Comment} $\sharp$\thecomment.~PP:}}%
{\end{quote}}

{\refstepcounter{comment}%
\begin{quote}
\color{\VBcolor}\small$\blacksquare$ \textbf{\underline{Comment} $\sharp$\thecomment.~VB:}}%
{\end{quote}}

{\refstepcounter{comment}%
\begin{quote}
\color{\JScolor}\small$\blacksquare$ \textbf{\underline{Comment} $\sharp$\thecomment.~JS:}}%
{\end{quote}}

{\refstepcounter{comment}%
\begin{quote}
\color{\FSBcolor}\small$\blacksquare$ \textbf{\underline{Comment} $\sharp$\thecomment.~FSB:}}%
{\end{quote}}

\definechangesauthor[color=\HMcolor]{HM}
\definechangesauthor[color=\CMcolor]{CM}
\definechangesauthor[color=\JMMCcolor]{JMMC}
\definechangesauthor[color=\PPcolor]{PP}
\definechangesauthor[color=\VBcolor]{VB}
\definechangesauthor[color=\JScolor]{JS}
\definechangesauthor[color=\FSBcolor]{FSB}
\definechangesauthor[color=\MDcolor]{MD}

\begin{document}


\title{Accessing the pion 3D structure at US and China Electron-Ion Colliders}



\author{José Manuel Morgado Chávez}
\email{josemanuel.morgado@dci.uhu.es}
\affiliation{Department of Integrated Sciences and Center for Advanced Studies in Physics, Mathematics and Computation, University of Huelva, E-21071 Huelva, Spain}

\author{Valerio Bertone}
\email{valerio.bertone@cea.fr}
\affiliation{Irfu, CEA, Université Paris-Saclay, 91191, Gif-sur-Yvette, France}

\author{Feliciano De Soto Borrero}
\email{fcsotbor@upo.es}
\affiliation{Dpto. Sistemas F\`isicos, Qu\`imicos y Naturales, Universidad Pablo de Olavide, E-41013 Sevilla, Spain}

\author{Maxime Defurne}
\email{maxime.defurne@cea.fr}
\author{Cédric Mezrag}
\email{cedric.mezrag@cea.fr}
\author{Hervé Moutarde}
\email{herve.moutarde@cea.fr}
\affiliation{Irfu, CEA, Université Paris-Saclay, 91191, Gif-sur-Yvette, France}

\author{José Rodríguez-Quintero}
\email{jose.rodriguez@dfaie.uhu.es}
\affiliation{Department of Integrated Sciences and Center for Advanced Studies in Physics, Mathematics and Computation, University of Huelva, E-21071 Huelva, Spain}

\author{Jorge Segovia}
\email{jsegovia@upo.es}
\affiliation{Dpto. Sistemas F\`isicos, Qu\`imicos y Naturales, Universidad Pablo de Olavide, E-41013 Sevilla, Spain}


\date{\today}

\begin{abstract}
We present in this letter the first systematic feasibility study of accessing
generalised parton distributions of the pion at an electron-ion collider through
deeply virtual Compton scattering at next-to-leading order. It relies on a
state-of-the-art model, able to fulfil by construction all the theoretical constraints imposed
on generalised parton distributions. Strikingly, our analysis shows that quarks and gluons interfere destructively and that gluon dominance could be spotted by a sign change of the DVCS beam spin asymmetry.
\end{abstract}


\maketitle

\section{Introduction\label{sec:Intro}}

Due to its double role of being both a Goldstone boson of the chiral symmetry
breaking and a QCD bound state, the pion has been widely investigated since its
discovery in 1947. In the 1980s and after, many efforts have been done to
extract experimental information about its internal structure, from its electromagnetic form
factor (EFF) through pion-electron scattering \cite{Amendolia:1986wj} to its
parton distribution functions (PDFs) through the Drell-Yan process
\cite{Conway:1989fs,Bordalo:1987cr,Betev:1985pf}.
The latter has in fact triggered a controversy on the PDF large-$x$ behaviour which remains not completely solved today, despite modern phenomenological and theoretical progresses \cite{Barry:2018ort,Novikov:2020snp,Barry:2021osv,Aicher:2010cb}.
However, using the available pion sources, EFF measurements are limited to low momentum transfer, precluding the possibility to test perturbative-QCD (pQCD) predictions
\cite{Efremov:1979qk,Lepage:1980fj}.
Exploiting the ideas of Sullivan \cite{Sullivan:1971kd}, consisting of interacting with the meson cloud of the proton, EFF data at significantly larger values of the momentum transfer between the virtual and real pion have been obtained \cite{Huber:2008id}.
This same principle  is
being seriously considered both for improving the knowledge on the pion EFF and
to extract the PDFs of the pion in the context of the forthcoming US and Chinese Electron
Ion Collider (EIC and EicC) \cite{Aguilar:2019teb,Arrington:2021biu,Anderle:2021wcy}.
The question has raised sufficient interest so that the EIC Yellow Report \cite{AbdulKhalek:2021gbh} mentions the study of the 3D structure of the pion through the Sullivan process. The present paper is a quantitative assessment of the latter.

The 3D structure \cite{Burkardt:2000za} of the pion can be gained through
Generalised Parton Distributions (GPDs) \cite{Mueller:1998fv,Ji:1996nm,Ji:1996ek, Radyushkin:1996ru,Radyushkin:1997ki}. Throughout the years, many models for pion's GPDs have been developed  \cite{Frederico:2009fk,Mezrag:2013mya,Mezrag:2014tva,Mezrag:2014jka,Mezrag:2016hnp,Fanelli:2016aqc,Rinaldi:2017roc,Chouika:2017dhe,Chouika:2017rzs,deTeramond:2018ecg,Shi:2020pqe,Zhang:2020ecj,Zhang:2021mtn}, including in the crossed channel \cite{Kumano:2017lhr}. They rely on various physics assumptions and if feasible, Deep Virtual Compton Scattering (DVCS) \cite{Ji:1996nm} off the pion would provide key constraints on these models \cite{Bertone:2021yyz,Bertone:2021wib,Chavez:2021llq}.

In this paper, relying on the implementation of the state-of-the-art model for
pion's GPD presented in section 2, we compute in section 3 the Sullivan
amplitude at Next-to-Leading Order (NLO), the minimal order required to treat
the EIC kinematical region. In section 4, we evaluate the associated counting
rate and assess the asymmetries, concluding that, providing that the one-pion exchange is the dominant process, DVCS off a virtual pion will be measurable and will provide
 a clear signal for a glue-dominated regime ideal to ``understand the glue that binds us all''.

\section{Modelling GPDs\label{sec:GPDs}}

Among all available pion's GPD models, we chose the one presented in Ref.~\cite{Chavez:2021llq}, a kinematical completion of that featured in \cite{Zhang:2021mtn} owing to a long effort developed over the last decade \cite{Mezrag:2013mya,Mezrag:2014tva,Mezrag:2014jka,Mezrag:2016hnp,Chouika:2017dhe,Chouika:2017rzs}. In a nutshell, this model is built on the state-of-the-art Continuum Schwinger Methods (CSM) investigations \cite{Binosi:2016rxz,Qin:2016fbu,Qin:2020jig,Ding:2019lwe,Ding:2019qlr,Cui:2020tdf,Cui:2020dlm} which have already provided the community with Parton Distributions Functions (PDFs) in agreement with the large-$x$ behaviour extracted from experimental data including soft-gluon (threshold) resummation \cite{Aicher:2010cb,Barry:2021osv}; and are confirmed both in the quark and gluon sectors \cite{Fan:2021bcr,Chang:2021utv} by lattice QCD computations.

The leap from quark PDFs $q_\pi$ to quark GPDs $H^q_\pi$ is however made difficult because their dependences on $x$, the average momentum fraction of the active parton in the pion, $\xi$, half of the exchanged longitudinal momentum fraction, and $t_\pi$ the square of the total momentum transfer are constrained by a set of properties \cite{Diehl:2003ny,Belitsky:2005qn}. To ensure that all properties are satisfied by construction, we turn to the lightfront wave function formalism \cite{Diehl:2000xz} (see also  \cite{Frederico:2011ws,Chang:2013pq,Carbonell:2017kqa,Salme:2017oge,Mezrag:2020iuo} for details of the connection between CSM and lightfront physics) supplemented by the so-called GPD covariant extension \cite{Chouika:2017dhe,Chouika:2017rzs}. Our quark GPD in the DGLAP ($|x|\ge |\xi|$) region reads\, \cite{Chavez:2021llq}:
\begin{align}
  \label{eq:ModellingMasterRelation}
  \left.H^{q}_{\pi}\left(x,\xi,t_\pi\right)\right|_{x\geq\left|\xi\right|}=\sqrt{q_{\pi}\left(x_{in}\right)q_{\pi}\left(x_{out}\right)}\Phi^{q}_{\pi}\left(x,\xi,t_\pi\right)
\end{align}
with
\begin{align}
  \label{eq:DSEPDF}
  & q_\pi(x) = \mathcal{N}_{q}x^{2}\left(1-x\right)^{2}\left[1+\gamma x\left(1-x\right)+\rho\sqrt{x\left(1-x\right)}\right], \\
  \label{eq:PhiT}
  &\Phi^{q}_{\pi}\left(x,\xi,t_\pi\right)=\frac{1}{4}\frac{1}{1+\zeta^2}\left(3+ \frac{1-2\zeta}{1+\zeta}\frac{\textrm{arctanh}\left(\sqrt{\frac{\zeta}{1+\zeta}} \right)}{\sqrt{\frac{\zeta}{1+\zeta}}} \right), \\
  \label{eq:xinout}
   & \zeta = -\frac{t_\pi}{4M^2}\frac{(1-x)^2}{1-\xi^2}, \quad x_{in}= \frac{x+\xi}{1+\xi}, \quad x_{out} = \frac{x-\xi}{1-\xi};
\end{align}
modeled at a low reference scale  $\mu_{\textrm{Ref}}=331\, \textrm{MeV}$, defined as discussed in\;\cite{Cui:2020dlm,Cui:2020tdf}.
The PDFs parameters obtained in \cite{Ding:2019lwe} are $\mathcal{N}_{q}=213$, $\gamma=2.29$ and $\rho=-2.93$.
The covariant extension provides us with the corresponding ERBL ($|x| \leq |\xi|$) region. Evolution equations from $\mu_{\textrm{Ref}}$ up to experimental scales are applied in the scheme introduced in \cite{Ding:2019lwe,Cui:2020tdf,Zhang:2021mtn}, generating accordingly the gluon GPDs, which are crucial ingredients in any EIC-related study. The latter yields in the forward limit a gluon PDF in excellent agreement with the one computed on the lattice \cite{Fan:2021bcr,Chang:2021utv}.
This theoretical framework is reinforced by information coming from experimental extractions of EFF \cite{Huber:2008id}, fixing $M=318\, \textrm{MeV}$.

In the case of the Sullivan process, one should consider a \emph{transition} GPD between a virtual and a real pion. However, for low enough virtuality, it has been shown \cite{Qin:2017lcd,Perry:2018kok} that extrapolations from the standard GPD are possible. Therefore, virtuality effects are neglected in the chosen GPD model.  This relies on a previous work on the pion EFF \cite{Qin:2017lcd}, indicating that neglecting virtualities tends to yield a smaller signal at non-vanishing momentum transfer, thus not undermining our conclusion on the feasibility of the measurement.

\section{DVCS through the Sullivan process \label{sec:sullivan}}

\begin{figure}[t]
  \centering
  \begin{tikzpicture}
    \begin{feynman}
      \vertex (a) {\(e^{-}(l)\)};
      \vertex [right=of a] (b);
      \vertex [above right=of b] (f1) {\(e^{-}(l')\)};
      \vertex [below right=of b] (c);
      \vertex [blob, below left=of c] (d) {\contour{white}{}};
      \vertex [left =of d] (e) {\(p\)};
      \vertex [below right= of d] (f) {\(n\)};
      \vertex [right =of c] (g);
      \vertex [above right =of g] (h) {\(\gamma~(q')\)};
      \vertex [below right =of g] (i) {\(\pi^{+}\)};
      \vertex (m) at (3.3,-1);
      \vertex (tp) at (3.5,-2.5);
      
      \diagram* {
        (a) -- [fermion] (b) -- [fermion] (f1),
        (b) -- [boson, edge label'=\(\gamma^{*}\left(q\right)\)] (c),
        (g) -- [boson] (h),
        (g) -- [scalar] (i),
        (d) -- [scalar] (c),
        (e) -- [fermion] (d),
        (d) -- [fermion] (f),
      };
    
    \end{feynman}
  
    \node[draw,fill=blue!10,ellipse, minimum height=1cm, minimum width=2.25cm,align=center, text width=1cm] at (m) {DVCS};
    \node at (tp) {$t_\pi$};
    \draw[thick] (2.3,-2) arc (-110:-70:3.05cm);
    \node at (1.25,-1.55) {$\pi^+ ~(p_\pi)$};
    \node at (0.8,-3.2) {$t$};
    \draw[thick] (0.65,-2.6) arc (190:300:0.54cm);
  \end{tikzpicture}
  \caption{DVCS off a virtual pion through the Sullivan process: The virtual pion from the proton absorbs the virtual photon $\gamma^*$ and subsequently emits a real photon $\gamma$ while turning itself real. The kinematical variables are defined as follows: $Q^2=-q^2$, $y=\frac{p\cdot q}{p\cdot l}$, $y_\pi=\frac{p_\pi\cdot q}{p_\pi\cdot l}$ and $x_B^{\pi}$=$\frac{Q^2}{2p_\pi\cdot q}$.}
  \label{fig:SullivanDVCS}
\end{figure}
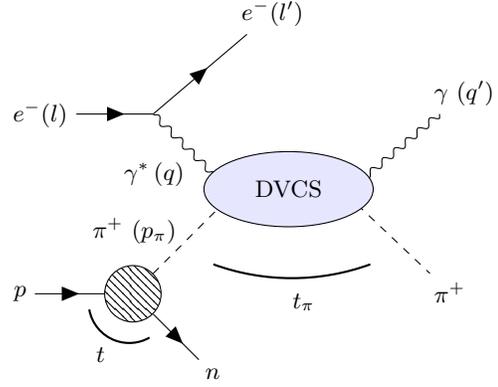

The DVCS amplitude is parametrised by a Compton Form Factor (CFF), a convolution of the pion GPD and a hard kernel computable within pQCD. 
At leading order (LO), the latter selects only quark GPDs, while at NLO, an exchange of a gluon pair in $t$-channel is allowed \cite{Ji:1997nk,Belitsky:1999sg,Pire:2011st,Moutarde:2013qs}.
Another competing process, called Bethe-Heitler (BH), leads to the same final state as DVCS except that the real photon is emitted by either the incoming or outgoing lepton. As a consequence the cross section for photon electroproduction off a pion $e \pi \rightarrow e \pi \gamma$ is the coherent sum of both processes amplitudes:

\begin{align}
\frac{d^5\sigma^{e\pi\to e\gamma\pi}(\lambda,\pm e)}{dy_{\pi} dx^{\pi}_B dt_{\pi} d\phi d\phi_e} &=\frac{d^2\sigma_0}{dQ^2 dx^{\pi}_B} \frac{1}{e^6} \nonumber \\
  & \times \left[ \left|\mathcal T^{BH} \right|^2 +  \left| \mathcal T^{DVCS}\right|^2  \mp
        \mathcal I\left(\lambda\right)  \right], \label{eq::xs}\\
 \frac{d^2\sigma_0}{dQ^2 dx^{\pi}_B} &=
\frac{\alpha_{\rm QED}^3 x^{\pi}_B y_{\pi}}{16\pi^2 Q^2 \sqrt{1+\epsilon^2}}, \\
\epsilon^2 &= 4 m_{\pi}^2 (x^{\pi}_B)^2/Q^2 , \\
\mathcal{I}\left(\lambda\right) & =\mathcal{I}_{\text{unp}}+\lambda\mathcal{I}_{\text{pol}};
\label{eq:dsig0}
\end{align}
where $\mathcal{T}^{BH}$ stands for the BH amplitude, $\mathcal{T}^{DVCS}$ for the DVCS amplitude and $\mathcal{I}_{\text{unp/pol}}$ for the un-/polarized interference terms. $\lambda$ is the electron helicity, $-e$ its charge, $\alpha_{QED}$ the electromagnetic coupling, $\phi_e$ the azimuthal angle of the scattered lepton, $m_{\pi}$ the mass of
the pion and $\phi$ the angle between the leptonic plane and the hadronic plane,
defined according to the Trento convention \cite{Bacchetta:2004jz}.
While the squared DVCS amplitude gives access to the
modulus of the CFF, its real and imaginary parts are independently accessible thanks to the interference term. In order to compute CFFs from the GPD model introduced above, we took advantage of \texttt{PARTONS} \cite{Berthou:2015oaw} where the formulae of Refs. \cite{Pire:2011st,Moutarde:2013qs} are implemented, together with \texttt{Apfel++} \cite{Bertone:2013vaa,Bertone:2017gds,EvolutionPaper} providing us with a LO evolution routine for GPDs.

Two additional variables are required to characterise the virtual pion in the
initial state: $x_{\pi}=\frac{p_{\pi}\cdot l}{p \cdot l}$ being the fraction of
the proton energy carried by the virtual pion in the $ep$ center-of-mass frame
and $t=p_\pi^2$. Following the formulae in Ref.~\cite{Amrath:2008vx}, the cross
section of the Sullivan process (see Fig. \ref{fig:SullivanDVCS}) then reads:
\begin{multline}
\frac{d^8\sigma^{\textrm{Sul}}(\lambda,\pm e)}{dy dQ{^2} dt_{\pi} d\phi d\phi_e dt d x_{\pi} d\phi_n} =\\
 x_{\pi} \frac{g^2_{\pi NN}}{16 \pi^3} F(t)^2 \frac{-t}{(m_{\pi}^2-t)^2} \;  |J^{Q^2}_{x^{\pi}_B}| \; 
 \frac{d^5\sigma^{e\pi\to e\gamma\pi}(\lambda, \pm e)}{dy_{\pi} dx^{\pi}_B dt_{\pi} d\phi d\phi_e} \label{eq::fullxs}
\end{multline}
with $J^{Q^2}_{x^{\pi}_B}$ the Jacobian between $Q^2$ and $x^{\pi}_B$, $\phi_n$ the azimuthal angle of the spectator neutron and $g_{\pi NN}=13.05$ being the pion-nucleon coupling constant. Amrath \emph{et al.} introduced a phenomenological factor $F(t)$ in Ref. \cite{Amrath:2008vx} to reduce the pion-nucleon vertex as $|t|$ increases. The latter takes the form:
\begin{equation}
  F(t)=\frac{\Lambda^2-m_{\pi}^2}{\Lambda^2-t},
  \label{eq::ref}
\end{equation}
with $\Lambda=800$~MeV. Finally, we stick to a chiral limit description of the
pion, neglecting its mass in Eq.~\eqref{eq::xs} and associated equations from
Ref.~\cite{Belitsky:2008bz}.

In order to ensure a proper interpretation of the event as a Sullivan DVCS
process, the pion virtuality $|t|$ must be small enough and will be taken such
that $|t| \le |t|_{max}= 0.6~\textrm{GeV}^2$ as well as
$s_{\pi}=\left(p_{\pi}+q\right)^2>s_{\pi}^{min}=4$~GeV$^2$ \cite{Amrath:2008vx}.
In addition, to reduce the contribution from nucleon resonances $N^*$ through
the process $ep\rightarrow eN\gamma \rightarrow en\pi \gamma$, the invariant
mass of the $n\pi$-system is required to be larger than 2~GeV. For the sake of completeness, let us mention that there also
are contributions which cannot be easily
restricted by cutting on kinematical variables.
For instance, virtual $\rho$'s may contribute as well through $\gamma^* \rho \rightarrow \gamma \pi$ and can only be assessed with a model of transition form factors and GPDs. We neglect them in the present study.

\section{Evaluation of observables\label{sec:Results}}

\begin{figure}[t]
  \centering
  \includegraphics[width=9cm, height = 5cm]{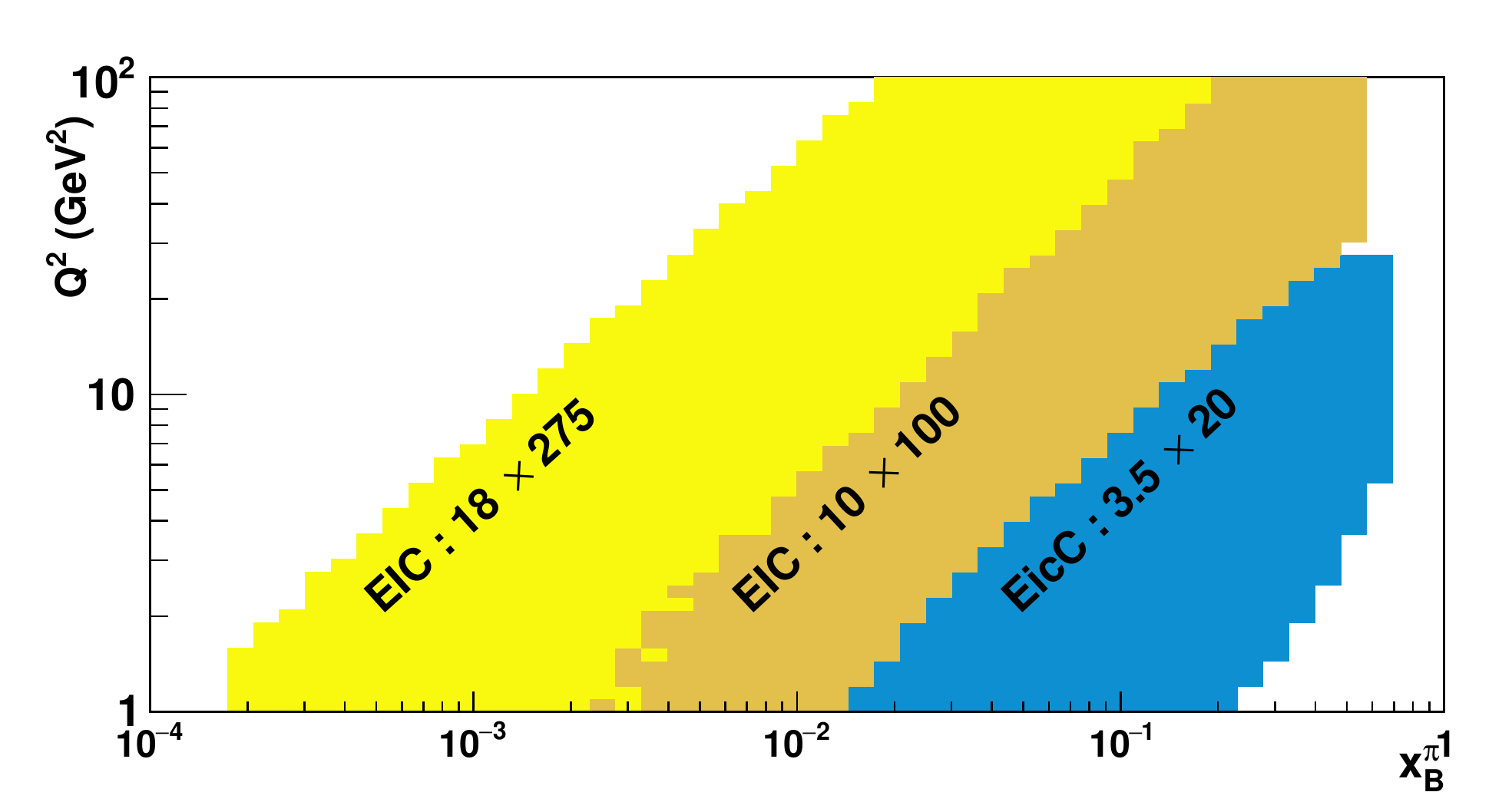}
  \caption{Phase space in $x^{\pi}_B$ and $Q^2$ considered in the present study: Facilities and configurations (electrons $\times$ protons beams energies in GeV) contributing the most to the statistics in the colored areas are specified.}
  \label{fig:PhaseSpace}
\end{figure}

Both EIC and EicC facilities will deliver highly polarised
lepton and hadron beams.
Their characteristics are summarised in Table~\ref{tab::EiC} and the coverage
they offer is illustrated on Fig. \ref{fig:PhaseSpace}.
\begin{table}[t]
  \begin{tabular}{|c||c|c|}
     \hline
    & EIC & EicC \\
    \hline \hline
    Lepton beam energy (GeV) & 5/10/18 & 3.5 \\
     \hline
     Hadron beam energy (GeV) & 41/100/275 & 20 \\
      \hline
      Lepton polarization & 70\% & 80\% \\
       \hline
       Hadron polarization & 70\% & 70\%  \\
        \hline
        Integrated luminosity (fb$^{-1}$/year) & 10 & 50 \\
        \hline
         
  \end{tabular}
  \caption{Main characteristics of both electron-ion colliders obtained from Ref. \cite{AbdulKhalek:2021gbh} (EIC) and Ref. \cite{Anderle:2021wcy} (EicC).}
  \label{tab::EiC}
\end{table}

Regarding the EIC's design~\cite{AbdulKhalek:2021gbh}, a central barrel detector, with two end-caps sitting in a $3 \text{T}$ solenoidal magnetic field, will be in charge of spotting the scattered lepton, the photon and the recoil pion with pseudo-rapidity between $-4$ to $4$.
 For the purpose of this study, it is worth mentioning that a far-forward spectrometer will detect the recoil pion with polar angle between 6 and 20 mrad. A Zero-Degree calorimeter will detect the neutron with polar angles from 0 to 5.5 mrad. 

Equivalently, the design of the EicC~\cite{Anderle:2021wcy} is composed of a central and far-forward detector package. However, as the EicC is still at a conceptual stage, key parameters for our study are not provided such as the acceptance for the neutron. Hence, we will assume an ideal geometry for the EicC spectator neutron tagger.

To guarantee the exclusivity, a reliable detection/identification of the electron, the photon and the neutron is assumed. Momentum conservation will be required and therefore pion identification is not considered as mandatory.

\begin{figure*}[t]
  \centering
  \includegraphics[width= 17cm, height= 8cm ]{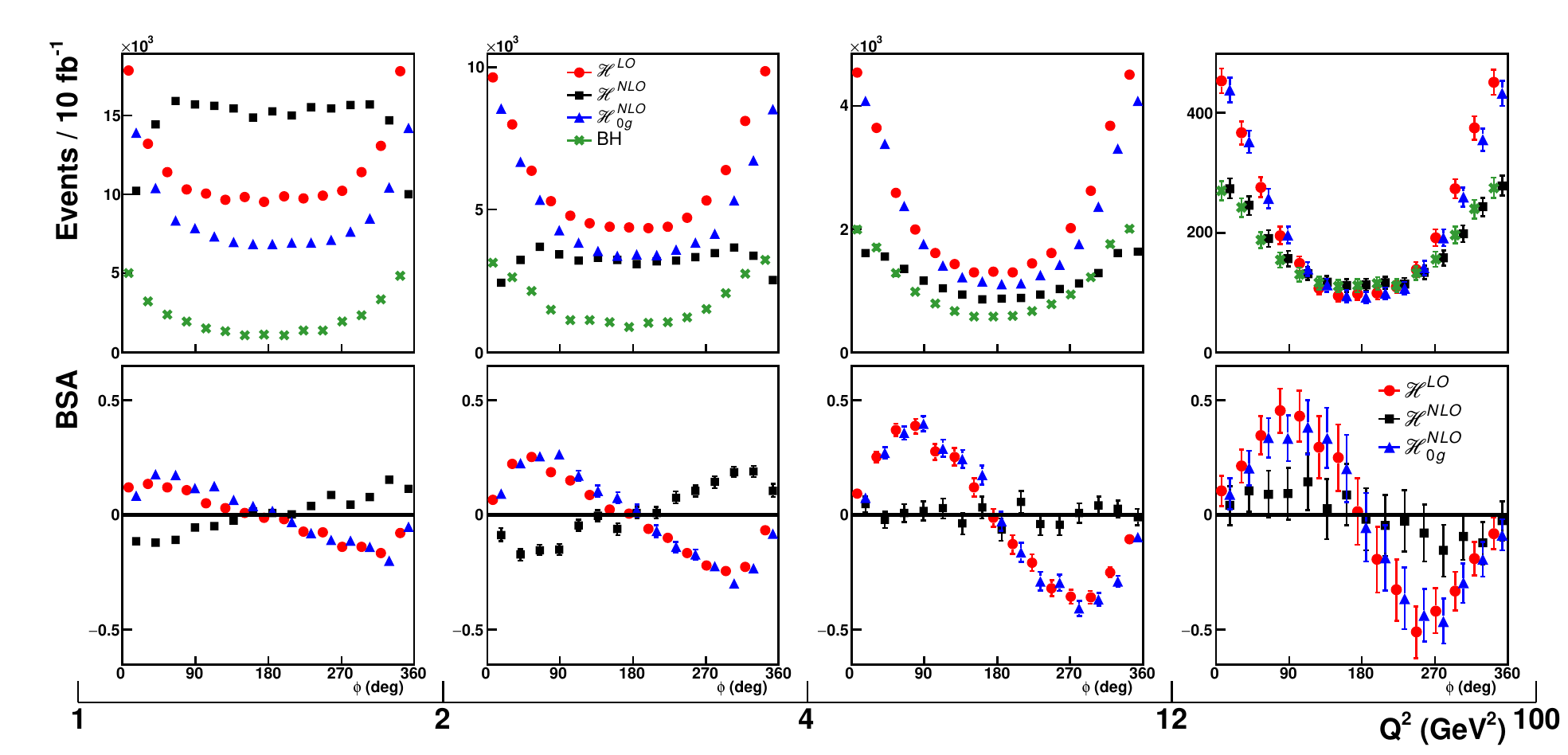}
  \caption{Number of DVCS events (upper charts) and expected beam-spin asymmetries (lower chart) as a function of $Q^2$ for $x^{\pi}_B\in [10^{-3};10^{-2}]$. Red circles: LO evaluation of the CFF; blue triangles: NLO evalution but without taking gluon GPDs into account; black circles: full NLO results. The BH event rates is as well displayed by the green crosses.}
  \label{fig:q2scan}
\end{figure*}    

The number of events is then estimated by Monte-Carlo simulation.
As seen in
Eq.~\eqref{eq::fullxs}, five kinematical variables and three angles are necessary
to fully determine the final state.
They are all uniformly generated in a range whose boundaries are controlled by previously generated variables, together with cuts guaranteeing the validity of the Sullivan process following Ref.~\cite{Amrath:2008vx}.
For each event, the generation starts with $x_{\pi}$ and
$t$ to determine the virtual pion. Lower and upper bounds for $x_{\pi}$ are defined as follows:
\begin{align}
  \label{eq:xpimin}
  x_{\pi}^{min} &= \frac{1}{y_{max}}\frac{s_{\pi}^{min}+Q^2_{min}}{s}, \\
  \label{eq:xpimax}
  x_{\pi}^{max} &= \frac{1}{2} \left[\sqrt{\tau^2+4\tau}-\tau\right],
\end{align}
with $\tau = |t|_{max}/M_N^2$.
Then $t$ is generated between $-|t|_{max}$ and $-|t|_{min}=\frac{x_\pi^2 M_N^2}{1-x_\pi}$, with $M_N$ being the nucleon mass. To complete the determination of the virtual
pion and neutron spectator, a rotation of an angle $\phi_N \in \left[0 ; 2\pi\right]$ around the incoming proton is performed. Then, the values of  $Q^2$ and $y$ are generated, providing us with the virtual photon. Considering that cross sections are too low above $Q^2=100$~GeV$^2$, $Q^2_{max}$ is given by:
\begin{equation}
Q^{2}_{max}=\min\left(sx_{\pi}^{max}y_{max}-s_{\pi}^{min},100\right)\;,
\end{equation}
with $s=\left(p+l\right)^2$. The upper
bound for the inelasticity $y$ is set to 0.85 \cite{Amrath:2008vx}, while the
lower bound is given by:
\begin{equation}
y_{min}=\frac{1}{x_{\pi}}\frac{s_{\pi}^{min}+Q^2_{min}}{s}\;.
\end{equation}
Then the virtual photon and the scattered lepton are rotated around the incoming lepton by the angle $\phi_e \in \left[0 ; 2\pi\right]$.

Finally, the variable $t_{\pi}$ lies within the range
$t_{\pi}^{max}=-0.6$~GeV$^2$ and $t_{\pi}^{min}$
computed exactly, \emph{i.e.} taking into account the virtuality of the initial
state pion. A last rotation around the virtual photon axis in the $\gamma^*
\pi$-system is performed for $\phi$. In order to correct for the non-uniformity
of the event generation across the entire phase space, each event $i$ must be
weighted by a phase space factor $d\Psi^i$ defined as:
\begin{equation}
  d\Psi^i=\frac{1}{N_{gen}}\displaystyle \prod_{k \in \mathcal{K}} k^i_{max}-k^i_{min}\;,
\end{equation}
with $\mathcal{K}= \left\{Q^2, y, t, x_\pi, t_\pi, \phi_e, \phi_N, \phi \right\}$ and $N_{gen}$ the number of generated events. The number of expected events $\mathcal{N}$ is then obtained by:
\begin{equation}
\mathcal{N}=\mathcal{L} \sum_{i \in \Phi} \frac{d^8\sigma^i(\lambda,\pm e)}{dy dQ{^2} dt_{\pi} d\phi d\phi_e dt d x_{\pi} d\phi_n} \times d\Psi^i\;,
\end{equation}
where $\Phi$ is the phase space of events passing kinematical cuts with all final state particles detected, and $\mathcal{L}$ the integrated luminosity over a year (see Tab.~\ref{tab::EiC}).

\begin{figure}[t]
  \centering
  \includegraphics[width=\linewidth]{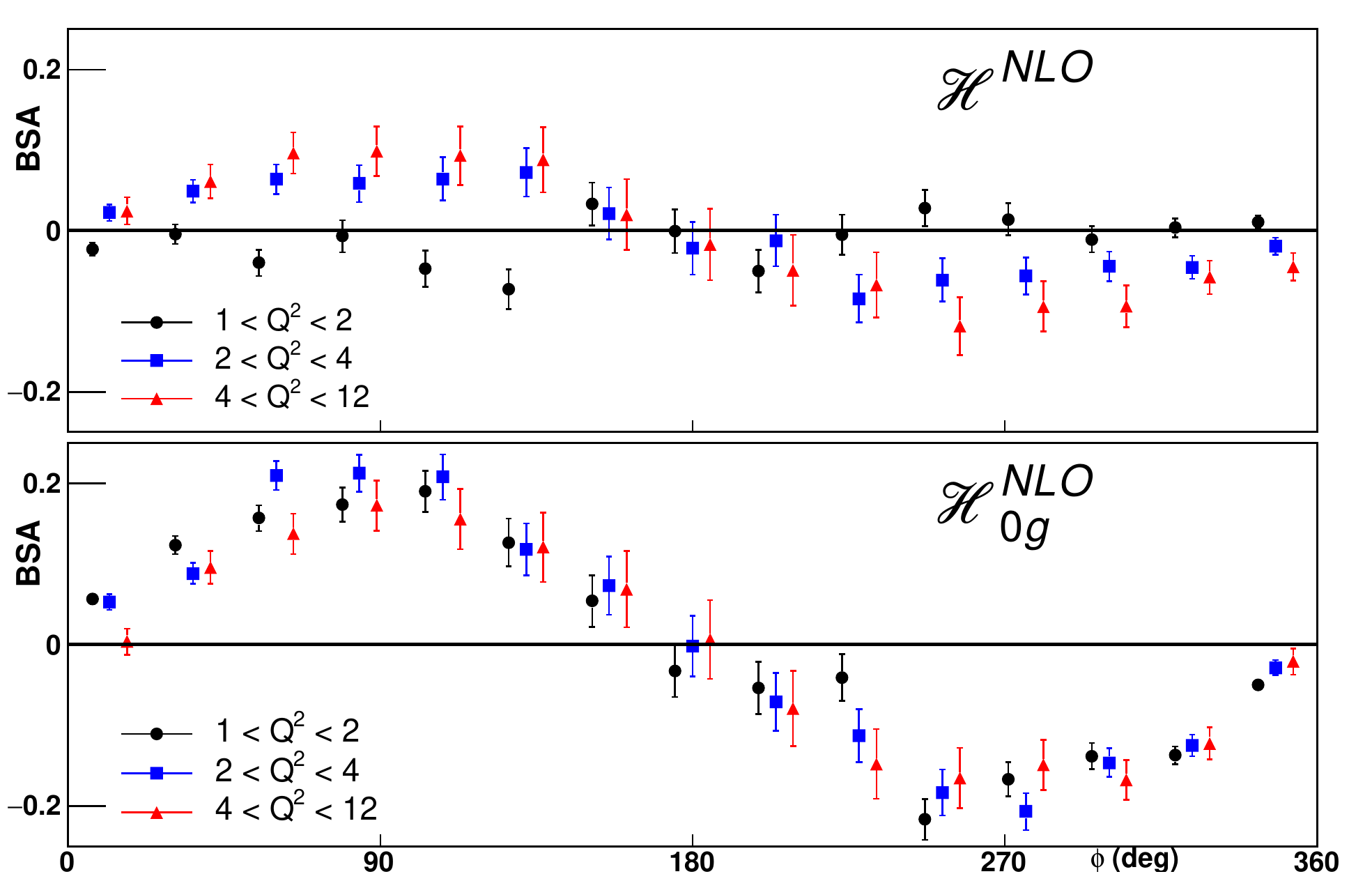}
  \caption{Expected beam-spin asymmetries as function of $\phi$ with $\mathcal{H}^{NLO}$ (top) and $\mathcal{H}^{NLO}_{0g}$ (bottom) from EicC for $x^{\pi}_B\in [0.1;0.5]$ and three different $Q^{2}$-bins: black circles for $Q^{2}$ between 1 and 2~GeV$^2$, blue squares between 2 and 4~GeV$^2$, and red triangles between 4 and 12~GeV$^2$.}
  \label{fig:q2scan_EICC}
\end{figure}

Three scenarii are compared with CFF-$\mathcal{H}$ computed at LO ($\mathcal{H}^{LO}$), at NLO  with ($\mathcal{H}^{NLO}$) and without ($\mathcal{H}^{NLO}_{0g}$) the contribution of the gluon GPD.
From Fig.~\ref{fig:q2scan} presenting expected count rates and Beam Spin
Asymmetries (BSA) for $x^{\pi}_B\in [10^{-3};10^{-2}]$  and 4 different $Q^2$-bins, several conclusions can be drawn.
First, we obtain event-rates clearly not compatible with the sole BH contribution,  highlighting the possiblity of accessing DVCS on a pion target at future electro-ion colliders.
Second, $\mathcal{H}^{LO}$ and $\mathcal{H}^{NLO}_{0g}$-predictions are similar, therefore the NLO corrections to the quark amplitude are small. On the contrary, the $\mathcal{H}^{NLO}$-CFF which includes gluons offers substantially different predictions compared to  $\mathcal{H}^{NLO}_{0g}$-CFF for both event rates and beam spin asymmetries. The gluon contribution to the CFF has an opposite sign to the quark one, resulting in a destructive interference. This destructive interference is clearly visible for Q$^2$ above 12~GeV$^2$, with the expected statistics dropping to the BH-only signal and the BSA being reduced by more than a factor 2. As Q$^2$ decreases, the gluonic contribution dramatically increases: the BSA changes sign and the expected statistics with gluons gets even much higher than the quark-only signal for $Q^2<2\, \text{GeV}^2$ despite the interference. Our model predicts that the BSA sign change will be clearly visible at the EIC, being a smoking gun for pinning down the transition between quark and gluon dominance in the DVCS description.

Remarkably, the gluon contribution remains sizeable
in the valence region accessible through EicC (see Fig.~\ref{fig:PhaseSpace})
with high statistical accuracy as shown in Fig.~\ref{fig:q2scan_EICC}.
 At NLO and without the
gluon contribution in the valence region, the BSA amplitude does not change much as a function of $Q^2$ and reaches about 0.2. Contrarily, when the gluon GPD is considered, it almost vanishes at low $Q^2$ and rapidly increases with $Q^2$ but remains smaller by a factor 2 compared to the gluonless case.

\section{Conclusion}

In this paper, we have studied the possibility to probe experimentally the pion
3D structure through the Sullivan process. Using a state-of-the-art model based
on Continuum Schwinger Methods, we obtained for one year of integrated
luminosity at EIC and EicC a significant number of events. It shows that DVCS off a virtual pion will be measurable, providing that the one-pion exchange is the dominant process.
Although focusing in this article on
Q$^2/x^{\pi}_B$-dependence, the statistics should be high enough to study the $t$-dependence as well.
We also highlighted a smoking gun for gluon dominance of the DVCS cross section; namely that the beam spin asymmetry changes sign because of destructive interferences between quarks and gluons. The wide kinematical coverage coupled with the high luminosity of EIC and EicC should allow us to see this effect.
Since the role of the two gluons exchange in the $t$-channel becomes dominant, next-to-next-to-leading order corrections to the DVCS kernel \cite{Braun:2020yib} are certainly desirable, and may confirm the behaviour highlighted here at NLO.

\begin{acknowledgments}
 We would like to thank P. Barry, H. Dutrieux, T. Meisgny, B. Pire, K. Raya, C.D. Roberts, Q.-T. Song, P. Sznajder and J. Wagner for interesting discussions and stimulating comments. This work is supported by University of Huelva under grant EPIT-2019 (J.M.M.C). F.S., J.R.Q. and J.S. aknowledge support from Ministerio de Ciencia e Innovación (Spain) under grant PID2019-107844GB-C22; Junta de Andalucía, under contract No. operativo FEDER Andalucía 2014-2020 UHU-1264517 and P18-FR-5057 and PAIDI FQM-370. This project was supported by the European Union's Horizon 2020 research and innovation programme under grant agreement No 824093.  This work is supported in part in the framework of the GLUODYNAMICS project funded by the "P2IO LabEx (ANR-10-LABX-0038)" in the framework "Investissements d’Avenir" (ANR-11-IDEX-0003-01) managed by the Agence Nationale de la Recherche (ANR), France.
\end{acknowledgments}

\bibliography{Bibliography}
\bibliographystyle{unsrt}

\end{document}